%Paper: cond-mat/9408035
%From: Klaus Ziegler <ziegler@tkm.physik.uni-karlsruhe.de>
%Date: Wed, 10 Aug 94 19:23:12 +0200

\magnification\magstep1
\baselineskip=16pt
\hsize = 6.9 true in
\language=0
\def\no{\noindent}
\def\a{\alpha}
\def\arctan{\rm arctan}
\def\se{s_z}
\def\s{\sigma}
\def\dm{\delta m}
\def\Tr{{\rm Tr}}
\def\bm{M'}
\def\be{\eta+\omega}
\def\om{\omega}
\def\dm{\delta M}
\def\bPhi{{\bar \Phi}}

\font\smallfont=cmsy5 at 0.002truept
\font\smmallfont=cmmi5 at 0.002truept

\no
%{\hskip10truecm}iqhe.tex/\today
\centerline{\bf Integer Quantum Hall Effect for Lattice Fermions}
\vskip 0.2in
\centerline{K. Ziegler}
\vskip 0.3in
\centerline{Institut f\"ur Theorie der Kondensierten Materie,
 Universit\"at Karlsruhe,}
\centerline{ Physikhochhaus, D-76128 Karlsruhe, Germany}
\bigskip
%\centerline{(\today)}
\vskip 0.6in
\noindent Abstract:\par
\no
\no
A two-dimensional lattice model for non-interacting fermions in a
magnetic field with
half a flux quantum per plaquette and $N$ levels per site is
considered. This is a model which exhibits the Integer Quantum Hall
Effect (IQHE) in the presence of disorder. It presents an alternative to
the continuous picture for the IQHE with Landau levels. The
large $N$ limit can be solved: two Hall transitions appear and there
is an interpolating behavior between the two Hall plateaux.
Although this approach to the IQHE is different from the traditional one
with Landau levels because of different symmetries (continuous for
Landau levels and discrete here), some characteristic features are
reproduced. For instance, the slope of the Hall conductivity is infinite at
the transition points and the electronic states are delocalized
only at the transitions.
\vfill
\eject
\no
We consider in two dimensions non-interacting lattice fermions in a
homogeneous magnetic field
with half a flux quantum per plaquette. This problem was originally studied
some time ago by Fisher and Fradkin [1]. It was shown by these authors that
the large scale limit is equivalent to a two-dimensional Dirac theory.
In a recent article Ludwig et al.
[2] extended this model by introducing a staggered chemical potential.
In the large scale limit this parameter appears as the mass $M$ of the Dirac
fermions. Therefore, the purpose of this parameter is to create a gap of the
effective Dirac theory between the particle and the hole band. Thus
two parameters control this system of non-interacting fermions: the
staggered chemical potential
and the energy of the particle $E$. In order to observe the integer
quantum Hall effect (IQHE) one must introduce disorder. In the traditional
approach to the IQHE, which is based on a continuum model with Landau levels
[3,4], this is achieved by using random fluctuations
of the chemical potential. However, for the model under consideration it
was argued by Ludwig et al. [2] that randomness of the chemical potential
alone cannot create a non-vanishing density of states
(DOS) for $M=E=0$. Their argument is based on
a perturbation theory w.r.t. disorder. Consequently, disorder in $M$ would
not lead to a physical Hall transition because of the absence of states at
the transition point $M=E=0$.
In order to get a non-zero DOS for the physical Hall system Ludwig et al.
introduced another random quantity, an additional potential $V$ which
corresponds to energy fluctuation.
The need of this potential is somehow surprising because a random
chemical potential was sufficient to get the physical Hall transition in
the famous approach to the IQHE by Pruisken et al. [3,4]. It will be argued
subsequentially that, in contrast to the claim by Ludwig et al.,
a random mass is also sufficient for a non-zero DOS in the lattice model
of Refs. [1,2].

First of all, we notice that the result of Ludwig et al. is in contradiction
to an earlier work [5] where it was
shown that the DOS is {\it non-zero} at $M=E=0$. This result is based on a
rigorous proof. Therefore, it would be interesting to understand why the
perturbative approach of Ludwig et al. does not reproduce the correct
behavior of the DOS.

We will explain in this article that disorder in the Dirac mass $M$ can
lead to fluctuations in $V$ (around $V=0$). This is due to the fact that
the same average Green's function can be created by different types of
external fluctuations.
Using $N$-level fermions we obtain the fluctuations in $V$ from a suitable
representation and a non-zero DOS in the large $N$ limit.
\bigskip
\no
To introduce the details of the model the pure system of non-interacting
fermions on a two-dimensional lattice is considered along the lines of
Refs.[1,2]. This describes lattice fermions with nearest neighbor hopping
rate $t=1$ and next nearest neighbor hopping rate $t'/4$ in a
staggered potential $\mu(-1)^{x+y}$. If we identify fermions with the four
corners of the unit cell the related Hamiltonian reads in Fourier
representation
$$H(k)=$$
$$
{\smallfont\smmallfont\thinmuskip=0.01mu
\pmatrix{
\mu&\!\!\!\!{1+e^{-ik_x}}&\tau(1-e^{-ik_y})(1-e^{-ik_x})&1+e^{-ik_y}\cr
1+e^{ik_x}&-\mu&1+e^{-ik_y}&-\tau(1-e^{-ik_y})(1-e^{ik_x})\cr
-\tau(1-e^{ik_y})(1-e^{ik_x})&1+e^{ik_y}&\mu&-1-e^{ik_x}\cr
1+e^{ik_y}&\tau(1-e^{ik_y})(1-e^{-ik_x})&-1-e^{-ik_x}&-\mu\cr
}}$$
$$\eqno (1)$$
with $\tau=it'/4$.
Expansion of $k=(\pi,\pi)+ap$ for small $p$ vectors leads to the large
scale approximation which breaks up the Hamiltonian (1) into two independent
Dirac Hamiltonians
$H_\pm=\sigma\cdot p+\sigma_3(m\pm T')$ with Pauli matrices $\s_j$.
The lattice constant $a$ is scaled out with $T'=t'/a$ and $m=\mu/a$.
The two Dirac theories describe particles with different masses $m\pm T'$,
respectively. For large scale properties like the Hall transition it is
sufficient to consider only the light particle with $M=m-T'$.
The Hamiltonians obey individually a discrete symmetry $H_\pm\to -\s_3
H_\pm\s_3$ for zero mass [2,5]. This is an important observation because
the effective theory for the Landau level system obeys a continuous symmetry
[3,4]. That means the two approaches to the IQHE are fundamentally different.
Nevertheless, we will see that certain physical properties are similar
in both cases.

The dispersion of $H_-$ is $E=\pm\sqrt{M^2+p^2}$.
This is a Dirac theory with a particle and a hole band which are
separated by a gap with width $2M$.
We can vary the
number of particles/holes by varying the width of the gap.
Suppose, the energy is in the lower (hole) band. Varying the gap means
we add/remove particles and holes to/from the edges of the gap.
This has no effect
on the current because those states are not occupied. On the contrary,
we can have the energy in the upper (particle) band. Varying the gap
means that we add or remove the same number of particles and holes.
Consequently, there is again no net current change because both contributions
cancel each other. The situation is different if the energy is inside
the gap: only the lower (hole) band is completely filled whereas the
upper (particle) band is empty. Varying the gap means adding/removing
holes to/from the system if the energy passes a band edge. This implies an
additional current. Finally, holes and
particles can be exchanged by the transformation $M\to-M$. This gives a
particle (hole) current for $M>0$ ($M<0$), respectively. To make this more
explicit the Hall conductivity $\sigma_{xy}$ is calculated as the response
to an external static field $q_y$
$$\sigma_{xy}=j_x/E_y={i\over q_y}\int{\sum_{r,r'}}' Tr[\sigma_x
(H-i\omega+E)^{-1}_{r,r'}(H-i\omega+E+q_y\sigma_y)^{-1}_{r',r}]{d\omega
\over2\pi}\eqno (2)$$
where $H$ is either $H_-$ or $H_+$ and $\sum'$ is the sum normalized with
the number of lattice sites.
Using the Green's function $G(E-i\om)=(H-i\omega+E)^{-1}$ we obtain for
$q_y\sim0$
$$\sim i{\sum_{r,r',r''}}'\int Tr[\sigma_xG(E-i\om;r,r')G(E-i\om;r',r'')
\sigma_yG(E-i\om;r'',r)]{d\omega\over2\pi}.$$
For infinite cut-off of the $k$--integration (i.e. infinite bands) we find
in units of $e^2/\hbar$ [2]
$$\sigma_{xy}={M\over4\pi|M|}\Theta(|M|-|E|),\eqno (3)$$
where $M$ can be the light or the heavy mass.
This result reflects correctly the qualitative interpretation of the Hall
conductivity. The Hall conductivity for the original
lattice fermion problem is the sum of the Hall conductivities from
the light and the heavy mass, such that the total $\sigma_{xy}$ has a jump
from 0 to 1.
\bigskip

Now we introduce disorder through a random Dirac mass $M$.
Disorder creates a DOS inside the gap of the pure
system. This means that the bands are broadened such that their inner
tails can overlap. The gap is closed if the fluctuations
$\dm$ are larger than $M$ [5]. That means we have a compact region in $M$
around $M=0$ for which the DOS $\rho(E)$ is non-zero. We cannot
distinguish particle and hole contributions to the current as in the pure
Dirac theory because the particle hole symmetry is spontaneously broken in
this case [5]: due to the fluctuations in $M$ new complicated (mixed) states
are created inside the gap of the pure system. They lead to a new Hall
conductivity which interpolates between the two Hall plateaux. Once there
is a gap again (if $|M|$ is larger than the maximal fluctuations $|\dm|$)
then the Hall plateaux appear (Fig.1). In the following we will generalize
the model of Ref.[2] by introducing
$N$ levels for the fermion Hamiltonian $H_\pm$. Then it will be briefly
discussed that the large $N$ Hall conductivity describes indeed
such a behavior. Here we can use results obtained in previous studies of the
Dirac Hamiltonian $H_-$ [8,9].

\bigskip
\no
As a generalization of $H_\pm$ to $N$ levels of fermions we introduce
$H^{\alpha \alpha'}=H_0^{\alpha \alpha'}-\dm_{r}^{\alpha \alpha'}
\s_3$ ($\alpha ,\alpha'=1,2,...,N$)
with $H_0^{\alpha \alpha'}=H_\pm\delta^{\alpha \alpha'}$ and for the Hermitean
random matrix $\dm$ we assume
$$\langle\dm_{r}^{\alpha \alpha'}\dm_{r}^{\alpha'' \alpha'''}\rangle
={g\over N}\delta^{\alpha \alpha'''}\delta^{\alpha' \alpha''}.\eqno (4)$$
That means only random fluctuations couple the $N$ different Dirac systems.
Physical quantities like the DOS per site or the conductivities are
proportional to the number of levels $N$. It is natural to normalize these
quantities by $N$.

\no
The purpose of the $N\to\infty$ limit is to obtain in a simple way a compact
distribution to the DOS $\rho(E,M)=(1/2)\sum_\mu G_{\mu,\mu}(E+i0^+)$
from $\dm$ which fills the gap. This is a consequence of the
well-known fact that $N\to\infty$ gives a semicircular density $\rho(M)
\equiv\rho(E=0,M)$ [6]. However, for a non-compact
distribution (e.g., Gaussian) and finite $N$ there is no gap at all; the
non-compact fluctuations $\dm$ extend the DOS over the whole real axis.
The trick is now
to consider the large $N$ limit and $1/N$ expansion (a finite order of these
terms does not create particles outside the $N\to\infty$ DOS) as a
technical device in order to maintain a density $\rho(M)$ with compact
support. There are Hall transitions at the edges of this `band'
which was created by the mass fluctuations. This result indicates that the
Hall plateaux are only due to the opening of a
gap in $\rho(E,M)$. If we take $N<\infty$ there should be still a
qualitative change in $\sigma_{xy}$ due to a cross-over analogous to the
transition with the compact DOS because the DOS in the gap is exponentially
small for $|M|>M_c$. Therefore, $\sigma_{xy}$ reaches almost the plateau
value whereas we find a smooth change between the two plateau values by
varying $M$ between $-\infty$ and $\infty$.
Nevertheless, in a real system it is more natural to truncate the large
fluctuations.

\no
The Green's function can be expressed formally as a functional integral for
non-interacting fermions [7]
$$G_{\mu,\mu'}^{\a\a'}(z;r,r')=
[(H_0-\dm\s_3 +z\s_0)^{-1}]_{r,\mu;r',\mu'}^{\a\a'}
=-i\int{\bar \Psi}_{r',\mu'}^{\a'}\Psi_{r,\mu}^{\a}
\exp(-S_1)\prod_r d\Phi_r d{\bar \Phi_r}\eqno (5)$$
with the action (sum convention for $\a$)
$$S_1=i\se [-(\Phi,(H_0+z\s_0)\bPhi)+\sum_r\dm_{r}^{\a\a'}
(\Phi_{r}^{\a'}\cdot\s_3\bPhi_{r}^{\a})]\eqno (6)$$
with $\se ={\rm sign}({\rm Im}z)$ and the field
$\Phi_{r,\mu}^\a=(\Psi_{r,\mu}^\a,\chi_{r,\mu}^\a)$.
The first component is Grassmann and the second complex. The complex
component is added to normalize the functional integral in (5). Averaging
with Gaussian distributed fluctuations yields
$$S_2=-i\se(\Phi,(H_0+z\s_0)\bPhi)+ {g\over N}\sum_r(\Phi_{r}^\a\cdot
\s_3 \bPhi_{r}^\a)^2.\eqno (7)$$
Thus we have derived an effective field theory for $\Phi$ which serves as
a generating functional for the average Green's function.
It is important to notice that {\it not only} $\dm$ creates the
fermion-fermion
interaction in (7) but also other types of disorder. For instance, the
interaction can also be created by a term which couples to a matrix
field ($\mu=1,...,4$ includes the complex and Grassmann components):
$${N\over g}Q_{r;\mu,\mu'}(\s_3)_{\mu'}Q_{r;\mu',\mu}(\s_3)_{\mu}
-iQ_{r;\mu,\mu'}\Phi_{r,\mu'}^{\a}\bPhi_{r,\mu}^{\a}
\to {g\over N}\sum_r(\Phi_{r,\mu}^\a(\s_3)_\mu \bPhi_{r,\mu}^\a)^2
\eqno (8)$$
This implies that the distribution $\dm$ can be transformed into another
distribution with a new `random variable' $Q$ (which does not have a
probability measure but some generalized distribution including Grassmann
variables). In other words, we can write
$$\langle[(H_0-\dm\s_3 +z\s_0)^{-1}]^{\a\a}\rangle_{\dm}=
\langle[(H_0-Q+z\s_0)^{-1}]^{\a\a}\rangle_Q\eqno (9)$$
The distribution which belongs to $\langle ...\rangle_Q$ was investigated
in detail in [8,9]. Here we present just the result for leading order in
$N$:
$\langle ... \rangle_Q =\int ...\exp(-NS(Q,P))\prod_r dP_r dQ_r$
with complex fields $Q_r$, $P_r$ and
$$S(Q,P)={1\over g}\sum_r[\Tr_2(Q_r\s_3)^2+\Tr_2(P_r\s_3)^2]$$
$$+\log\det(H_0-2Q+z\s_0)-\log\det(H_0+2iP+z\s_0) +O(N^{-1}).\eqno (10)$$
The number of levels $N$ appears in front of the action.
Thus the effect of disorder for $N\to\infty$ can be evaluated in
saddle point approximation. The saddle point equation reads
$${\delta\over\delta Q}\big\lbrack {1\over g}\Tr (Q\s_3)^2+
\log\det(H_0-2Q+z\s_0)\big\rbrack=0.\eqno (11)$$
A second saddle point equation appears from the variation of $P$ by
replacing $Q\to -iP$.
As an ansatz we take a uniform saddle point solution
$Q_0 =-i P_0 =-(1/2)[i\eta\s_0+M_s\s_3]$.
Then (11) leads to the conditions
$\eta =(\be-iE) g I$, $M_s=-MgI/(1 +gI)$
with the integral
$I=\int\lbrack (M+M_s)^2 +(\be-iE)^2+k^2\rbrack^{-1}d^2k/2\pi^2$.
This means disorder shifts the frequency $\om\to\om+\eta$ and the Dirac
mass $M\to\bm=M+M_s$,
where $\eta(M,\om)$ and $M_s(M,\om)$ are solutions of the saddle point eqn.
The sign of $\eta$ is fixed by the condition that $\eta$ must be analytic in
$\om$. This leads to ${\rm sign}(\eta)={\rm sign}(\om)$.
Furthermore, $\rho(M)$ is proportional to $\eta$. The Hall conductivity
per fermion level reads
$$\s_{xy}={\bm\over2\pi}\int{1\over(\om+\eta-iE-i|\bm|)(\om+\eta-iE+
i|\bm|)}{d\om\over2\pi}\eqno (12)$$
and with the approximation that $\bm$ and $\eta$ do not depend on $\om$
we get
$$\s_{xy}\approx{1\over4\pi^2}[\arctan((\bm+E)/\eta)+{\rm arctan}
((\bm-E)/\eta)].\eqno (13)$$
The sum of $\s_{xy}$ for both masses are plotted in Fig.1 for $E=0$. An
analogous calculation gives for the longitudinal conductivity per level
$$\s_{xx}\approx{1\over4\pi^2}[\pi/2-\arctan((\eta^2+{M'}^2-E^2)
/2\eta |E|)].\eqno (14)$$
The transition between the Hall plateaux does not occur at $M=0$, as
suggested in Ref. [2], but at $M=\pm M_c$ where $\eta(M)$ vanishes. The
distance of these transitions is
small for weak disorder and $E=0$: $\sim\exp(-2\pi/g)$. Thus it seems to be
difficult to resolve the transitions in a real or numerical experiment.
The slope of the Hall conductivity $\s_{xy}(M)$
is infinite at the transition points in agreement with the Hall transitions
for the Landau levels [15].
In the pure limit ($\eta\to0$) the conductivity $\s_{xx}(M)$ is $1/4\pi$
in units of $e^2/\hbar$ outside and zero inside the gap. This is consistent
with the behavior of the Hall conductivity in the pure limit (eqn. (3)).
In the disordered system $\s_{xx}(M)$ vanishes also on the Hall plateaux
where $\eta=0$ provided $E^2<{M'}^2$. $\s_{xx}(M)$ vanishes always for $E=0$.
This reflects localization of the states which are created by disorder at the
center of the gap.
\bigskip
Up to now only properties which are obtained in the $N\to\infty$ limit were
considered. Of course this is not sufficient to evaluate
space dependent properties like the localization length. The latter
requires the fluctuations around the saddle point. This can be studied on
the basis of Ref.[9], where a divergent correlation length was found at the
transition points $M=\pm M_c$. For $M\ne\pm M_c$ all length scales are
finite which implies localization away from the transition points.
A detailed investigation of the localized states must include the
evaluation of the localization length exponent $\nu$. This exponent
is known for the continuous system with Landau levels
from semiclassical approximations [10] and numerical studies as
$\approx7/3$ [11-14]. However, other values were also found experimentally
as well as theoretically [4,13,15,16].

In conclusion, a discussion of the IQHE for lattice fermions with half a
flux quantum per plaquette and $N$ levels per site was presented. In the
large $N$ limit two Hall transition are found with an interpolating
behavior between two Hall plateaux in contrast to the single transition of
Refs.[2-4]. Although this system is not from the
same universality class as the one described in Refs.[3,4] for the Landau
level system, some properties are similar like the infinite slope of the
Hall conductivity at the transitions.
\vfill
\eject
References

[1] M.P.A.Fisher and E.Fradkin, Nucl.Phys.B251[FS13], 457 (1985)

[2] A.W.W.Ludwig, M.P.A.Fisher, R.Shankar and G.Grinstein, preprint
Princeton (1993)

{\ \ \ } [Phys.Rev.B (in press)]

[3] H.Levine, S.B.Libby and A.M.M.Pruisken, Phys.Rev.Lett. 51, 1915
(1983)

[4] H.Levine, S.B.Libby and A.M.M.Pruisken, Nucl.Phys. B 240 [FS12] 30,
49, 71 (1984)

{\ \ \ }A.M.M.Pruisken, Phys.Rev.Lett. {\bf 61}, 1297 (1988)

{\ \ \ }{\sl The Quantum Hall Effect}, edited by R.E.Prange and S.M.Girvin

{\ \ \ }(Springer-Verlag, New York, 1990)

[5] K.Ziegler, Nucl.Phys.B285[FS19], 606 (1987)

[6] M.L.Mehta {\sl Random matrices} (Acad. Press New York, 1967)

[7] J. W.Negele and H.Orland, {\sl Quantum Many - Particle Systems}

{\ \ \ }(Addison - Wesley, New York, 1988)

[8] K.Ziegler, Europhys.Lett.{\bf 14 } (5) (1991) 415

[9] K.Ziegler, Nucl.Phys.B344, 499 (1990)

%[10] F.Wegner, Z.Physik B38, 207 (1979)

[10] G.V.Mil'nikov and I.M.Sokolov, JETP Lett. {\bf 48}, 536 (1988)

%[11] Refs. [4-10] in Huckestein (PRL '94?)
[11] J.T.Chalker and P.D.Coddington, J.Phys.C {\bf 21}, 2665 (1988)

[12] B.Huckestein and B.Kramer, Phys.Rev.Lett. {\bf 64}, 1437 (1990)

[13] B.Mieck, Europhys.Lett. {\bf 13}, 453 (1990)

[14] Y.Huo and R.N.Bhatt, Phys.Rev.Lett. {\bf 68}, 1375 (1992)

[15]  H.P.Wei, D.C.Tsui, M.Paalanen, and A.M.M.Pruisken, Phys.Rev.Lett.
{\bf 61}, 1294 (1988)

[16] T.Ando and H.Aoki, J.Phys.Soc.Jpn. {\bf 54}, 2238 (1985)

\vfill
\eject
\no
Figure Caption
\bigskip
\no
Fig.1: Hall conductivity $\sigma_{xy}$ in units of $e^2/h$  as a function of
the staggered chemical potential $M$ for disorder strength $g=2.2$.

\bye